# On the Structure of Superconducting Order Parameter in High-Temperature Fe-Based Superconductors

*T.E. Kuzmicheva[1], A.V. Muratov[1], S.A. Kuzmichev[1,2], A.V. Sadakov[1], Yu.A. Aleshchenko[1,3], V.A. Vlasenko[1], V.P. Martovitsky[1], K.S. Pervakov[1,4], Yu.F. Eltsev[1], V.M. Pudalov[1,5]*

[1] P.N. Lebedev Physical Institute RAS, 119991 Moscow, Russia

[2] M.V. Lomonosov Moscow State University, 119991 Moscow, Russia

[3] National Research Nuclear University MEPhI (Moscow Engineering Physics Institute), 115409 Moscow, Russia

[4] International Laboratory of High Magnetic Fields and Low Temperatures, 53-412 Wroclaw, Poland

[5] National University "Higher School of Economics", 101000 Moscow, Russia

## 1. Introduction

Unconventional superconductivity in quasi-two-dimensional iron compounds discovered in 2008 [1] is undoubtedly a challenging issue of condensed matter physics. Demonstrating critical temperatures up to 56 K for bulk SmOFeAs [2], but up to 100 K in monolayer FeSe [3], and possessing huge critical fields as high as 200 T [4], pnictides and selenides seem to be rather hopeful for some applications [5,6]. Being antiferromagnetic metals with spin-density wave (SDW) ground state in stoichiometric composition, the majority of iron pnictides and selenides turn into superconductivity under electron or hole doping. However, the two exceptions, a well-known LiFeAs [7] and recently discovered ThNOFeAs similar to an oxypnictide (1111 family) and $T_C \approx 30$ K [8], are nonmagnetic in the stoichiometric state. Note the Fermi energy is rather low for iron-based superconductors. For example, for the "driving" band near the M point, the Fermi energy offset the bottom of the band is about 0.2 eV [9].

As compared to other iron superconductors, $BaFe_{2-x}Ni_xAs_2$ and $Ba_{1-x}K_xFe_2As_2$, representative members of the so-called Ba-122 family, attract the research community due to both moderate $T_C$ up to 38 K, and rather easy growth of qualitative and large single crystals with variable substitution. Band-structure calculations [10] showed several Fe 3d orbital bands crossing the Fermi level, form cylinder-like Fermi surface sheets significantly warped in out-of-plane direction, hole-like around the $\Gamma$ point, and electron-like around the M point. The presence of multiple superconducting condensates below $T_C$ and corresponding forbidden energy bands

(superconducting gaps) had unambiguous confirmation, both theoretical and experimental [11,12].

In order to explain the underlying pairing mechanism, three basic models were presented. One of them, s$^{++}$ model predicts a strong intraband pairing, and two competing pairings — via spin fluctuations and via orbital fluctuations enhanced by phonons [13,14]. This competition could lead to anisotropic (angle-dependent in a *k*-space) or even nodal order parameter with a strong intraband coupling [14]. In contrast, s± model is based on a pairing via spin fluctuations. Superconductivity is driven by interband interaction, herewith the wave functions of the two condensates are in antiphase, which formally leads to $\Delta_L \Delta_S < 0$, where $\Delta_L$ is the large gap, and $\Delta_S$ is the small gap. Nesting between the hole and electron Fermi surface sheets causes a magnetic resonance (a peak of dynamic spin susceptibility at the nesting vector and a certain energy) [15,16].

The model describing an establishment of spin and charge superstripes (nanoscale phase separation) predicts a drastic $T_C$ enhancement caused by a Feshbach-type resonance when a band edge is close to the Fermi level (Lifshitz transition). Besides, in this approach one should consider the Cooper pairs condensate in an intermediate regime, from Bardeen—Cooper—Schrieffer (BCS) condensation, typical for classical isotropic superconductors, to Bose—Einstein condensation, with hardly overlapping pairs [9]. This approach seems reasonable, when accounting the quasiclassic condition $w_D \gg 2\Delta$ ($w_D$ is the Debye energy) is violated for iron-based superconductors, and the average size of Cooper pairs is close to the overlapping distance.

Possible observation of flat bands in 1111 family reported in angle-resolved photoemission spectroscopy (ARPES) measurements [17] seemingly suppose a presence of extended van Hove singularities typical for quasi-two-dimensional compounds. In the case, the theory of cuprates by A.A. Abrikosov [18] could be used in order to explain the high $T_c$ and the $\Delta$ values as high as ~ 13 meV.

Each of these models suggests a certain structure of superconducting gap and a set of some other parameters of the material. Unfortunately, despite the intensive eight-year studies, the available experimental data are rather contradictory to make any conclusions about the pairing mechanism. No model has got an unambiguous experimental confirmation yet, thus making the main issues still unanswered. For example, the experimentally determined BCS ratio in Ba-122 family varies by a factor of six [19-33, for a review, see 12, 34]. A possible reason is the strong out-of-plane gap anisotropy in k-space which seems to "smear" the gap values obtained in bulk probes, and a sensitivity-to-surface of the superconducting properties. The gap temperature dependences measured in [28,29] using point-contact Andreev spectroscopy (PCAR) are typical for a strong interband coupling. A number of inelastic neutron scattering probes [35-38] reported rather sharp magnetic resonance peak in favor of the $s^\pm$ model [15,16]. Nonetheless, the experimentally observed magnetic resonance peak seems [13,14] not so pronounced as predicted for $s^\pm$, and the $s^\pm$ system would be unstable in a presence of impurities [13,39]. The absence of nesting evident from ARPES measurements [7] and a strong intraband coupling evaluated from direct temperature dependences of the gaps [34,40] also support $s^{++}$. On the other hand, a proximity to a Lifshitz transition, the larger gap developed in the smallest Fermi surface cylinder [7,17], and the observed nanoscale phase separation and Feshbach-type resonance [41,42] indirectly facilitate the superstripe model.

This brief review shows, the reliable experimental data are essential in order to reveal the pairing symmetry and further to find a way to enhance critical temperature. Here we present synthesis, characterization, and a comprehensive study of Ba-122 single crystals with various substitutions and $T_C$. In order to study the superconducting properties of Ba-122 compounds, we used five complementary techniques and obtained a comprehensive set of self-consistent data.

## 2. Synthesis and Characterization

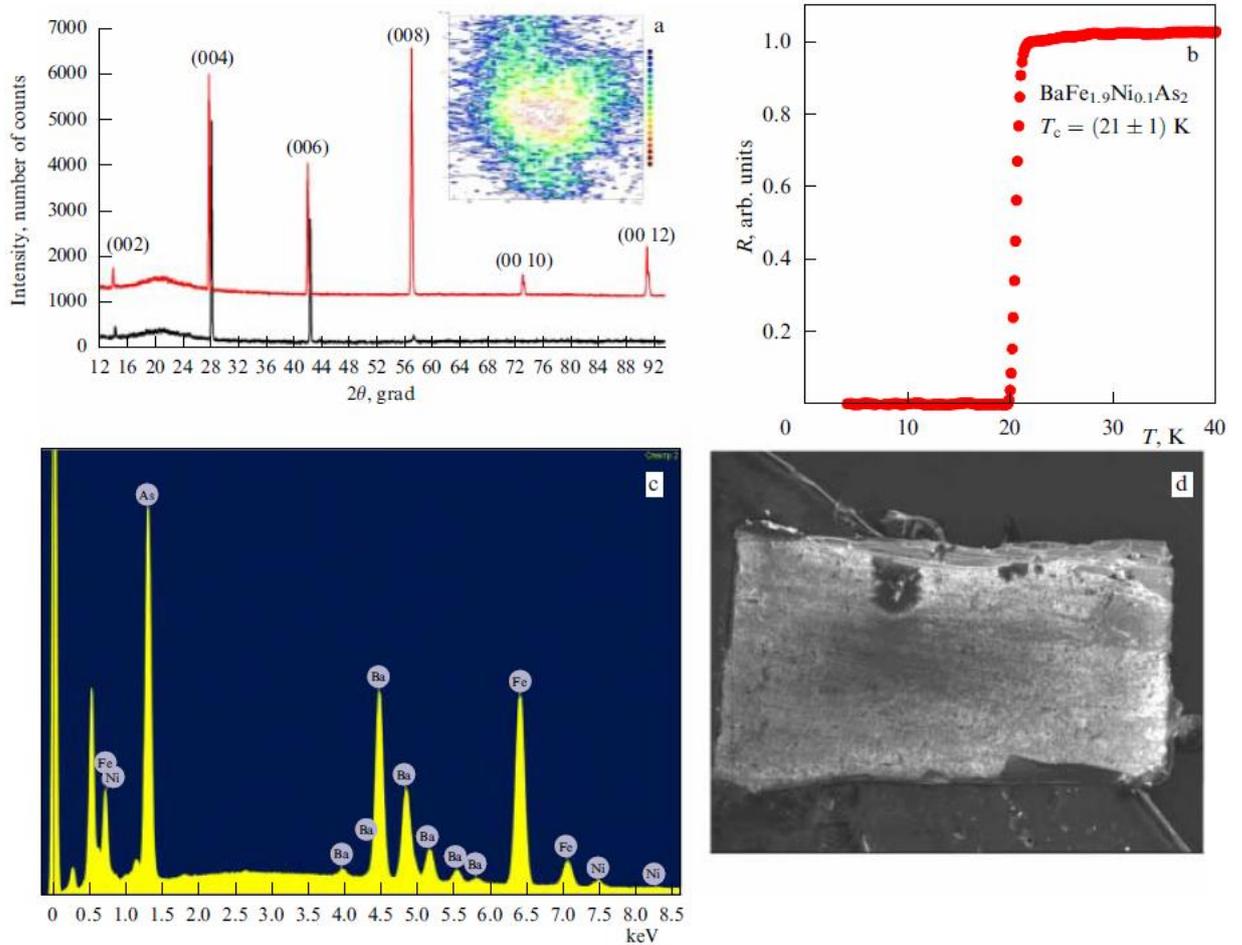

Fig. 1. a) X-ray diffraction spectra for $BaFe_{2-x}Ni_xAs_2$ single crystals with $x = 0.09$ (lower curve) and $x = 0.1$ (upper curve). The inset shows 2D-plot in the vicinity of (006) peak. b) Temperature dependence of resistance near the superconducting transition. c) Energy dispersive spectrum (EDS) of $BaFe_{1.92}Ni_{0.08}As_2$ single crystal. d) Electron microscope image of the $BaFe_{1.92}Ni_{0.08}As_2$ single crystal with dimensions 1.5·0.8 mm$^2$.

$BaFe_{2-x}Ni_xAs_2$ single crystals with various doping level and critical temperature up to $T_C$ = 20 K were synthesized using the self-flux technique. In order to prevent an oxidation in open air, all the reagents were weighted in a glove box with a controlled argon atmosphere. Metallic Ba, and high purity FeAs and NiAs precursors, preliminary obtained from the elements by solid-phase synthesis, were mixed in 1 : 5(1 − x) : 5x molar ratio, placed in the alumina crucible and sealed in a quartz tube under 0.2 bar argon pressure. The ampoule was heated up to 1200 °C and kept for 12 hours, trigging the reaction

$Ba + (2−x)FeAs + xNiAs = BaFe_{2-x}Ni_xAs_2$.

The long exposure time is necessary in order to complete the reaction with the development of the essential phase and to make the homogeneous flux, because the doping phase BaNi2As2 diffuse into the main phase BaFe2As2 via convection and is limited by a viscosity of the environment. After the exposure, the ampoule was cooled down to 1150 °C, with further flux crystallization in the temperature gradient, cooling it to 1050 °C with the rate 2°C per hour. When achieved the latter temperature, the liquid flux was decanted turning the ampoule.

The grown crystals were up to $4 \cdot 2 \cdot 0.2$ mm$^3$ in size. Figure 1a shows X-ray diffraction spectra of BaFe$_{2-x}$Ni$_x$As$_2$ single crystals with x = 0.09 (lower curve) and x = 0.1 (upper curve) obtained using the DRON-2.0 diffractometer with a curved graphite monochromator. The spectra shows intensive peaks attributed to the 122 phase barely, thus demonstrating the high purity of the synthesized single crystals. In order to detect twinning boundaries or disorientation of blocks in the single crystal, we measured the rocking curves and two-dimensional (2D) plots in the vicinity of the (006) peak using the diffractometer Panalytical X'Pert Pro MRD Extended. The fragment of XRD spectrum in the 2D vicinity of the (006) peak is detailed in the inset of Fig. 1a. Since the single peak in the vicinity of the (006) reflex is observed, thus evidencing the absence of disoriented blocks in the crystal. The single high-intensity peaks (Fig. 1a) evidence a high quality of the samples, and homogeneous Ni distribution within the bulk of the crystal. Energy dispersive spectroscopy (EDS) analysis showed the elemental ratio Ba:Fe:Ni:As ≈ 1.06:1.91:0.09:1.95 (Fig. 1c) agreeing well with the nominal ratio. Electron microscope image of BaFe$_{1.92}$Ni$_{0.08}$As$_2$ single crystal with dimensions $1.5 \cdot 0.8$ mm$^2$ is shown in Fig. 1d.

## 3. Study of the high field magnetization

The irreversible magnetization M(H,T) and magnetic susceptibility χ'(H,T) measurements were performed using the PPMS vibrating sample magnetometer in the fields up to 9 T applied along the crystal *c*-axis. The typical field sweep rate was 100 Oe/s. Resistive and magnetization measurements (Figure 2 a,b) showed clear and sharp superconducting transition with the presence of single superconducting phase and critical temperature $T_C \approx 19$ K of the $BaFe_{1.92}Ni_{0.08}As_2$ single crystal. With Ni concentration increase, the $T_C$ decreases. For the overdoped sample with x = 0.18, $T_C \approx 9.3$ K, the corresponding magnetic susceptibility transitions in various fields are shown in Fig. 2b. The superconducting transition width $\Delta T_C \approx 1.4$ K obtained using susceptibility (Fig. 2b) and resistance (Fig. 1b) temperature dependence, demonstrates structural perfection and homogeneity of the superconducting properties within the bulk.

Using magnetic hysteresis loop measurements, we plot the field dependence of critical current density $J_C$ for underdoped $BaFe_{1.92}Ni_{0.08}As_2$ (Figure 2c), which is similar to that for the samples with nearly optimal composition. The linear behaviour in I regime (untill 100–150 Oe at helium temperatures) is generally attributed with single vortex pinning mode. At higher fields up to 0.5 T, power law dependence $J_C \sim H^{-\alpha}$ is observed, thus the II regime corresponds to a significant increase in the number of vortices in the bulk and their interactions. The obtained index is $0.37 < \alpha < 0.43$, thus being a bit lower than the $\alpha = 5/8$ predicted theoretically for strong pinning centers. The observed discrepancy may point to a presence of some extended defects and scattered weak pinning centers [43]. The presence of the III regime with $J_C(H) \sim$ const seems to be caused by coexisting large and small pinning centers, acting like a cage for the magnetic vortices. The strong vortex pinning weaken in the IV regime is accompanied with the decreasing of $J_C$ and melting of vortex lattice. Fig. 2d shows the normalized pinning force $f_p = F_p/F_{pmax}$ as the function of normalized field $h = H/H_{irr}$ measured at various temperatures. The value of irreversible field $H_{irr}$ was determined as that corresponding to critical current density turning to zero ($J_c \rightarrow 0$). Obviously, the $f_p(h,T)$ curves merging at the field H || c. Using the Dew-Hughes model [44] with $f_p(h,T) \sim h^p(1 - h)^q$, we get the coefficients p = 1.64 and q = 3.43 for the $BaFe_{1.92}Ni_{0.08}As_2$ single crystal. In accordance with this model, the obtained peak $h^p = 0.32$ favors a prevalence of strong point pinning centers. Another evidence of the strong bulk pinning is a high symmetry of the magnetization loop at temperatures close to $T_c$, which also points the amount of magnetic impurities is insignificant in the single crystal (see the inset of Fig. 2d).

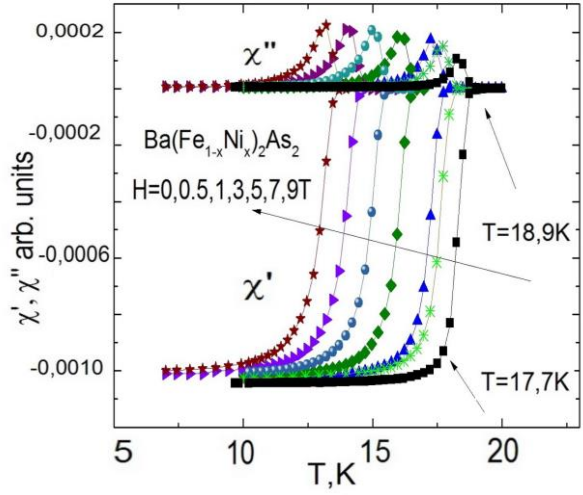
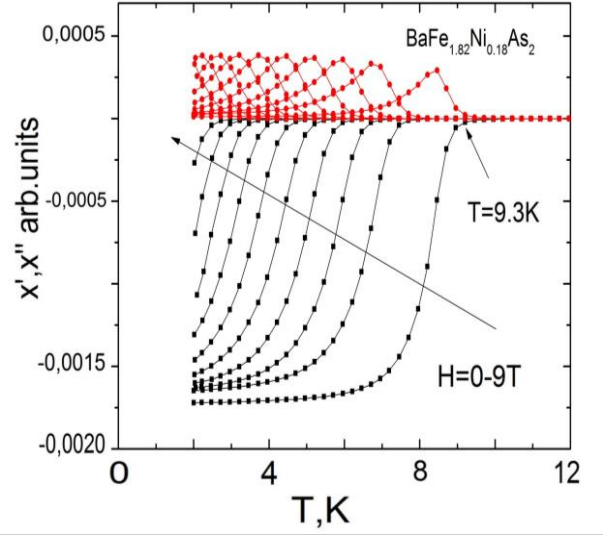
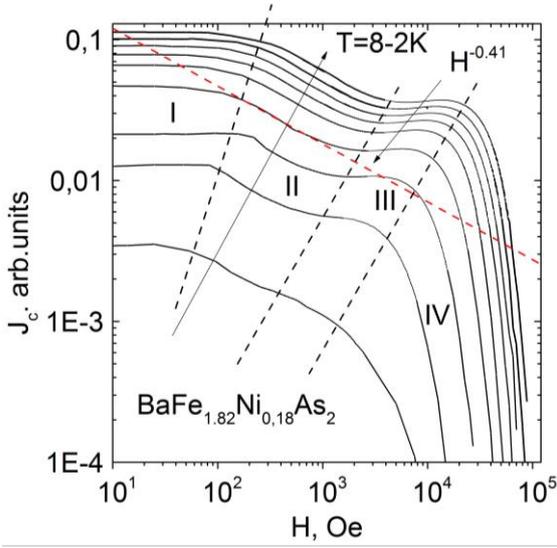
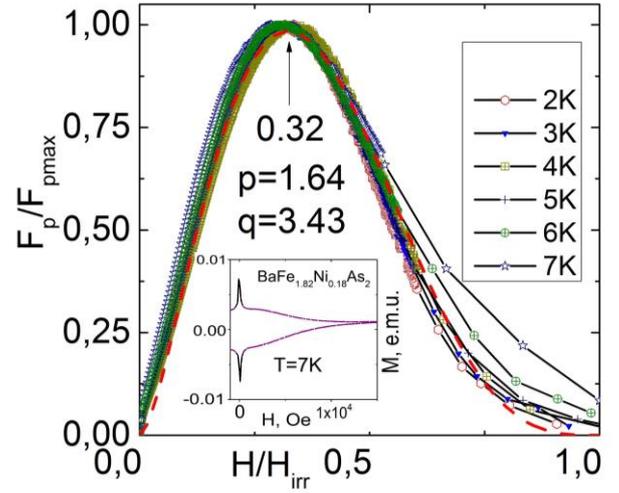

Fig. 2. Temperature dependence of magnetic susceptibility in various magnetic fields up to 9 T in (a) slightly underdoped $BaFe_{1.92}Ni_{0.08}As_2$ single crystal, and (b) in overdoped $BaFe_{1.82}Ni_{0.18}As_2$ with $T_c \approx 9.3$ K. c) Critical current density $J_C$ as function of magnetic field H, data are taken from [40]. d) Normalized pinning force $f_p = F_p/F_{pmax}$ as function of normalized field $h = H/H_{irr}$ measured at various temperatures. The inset shows the hysteretic loop at T = 7 K.

## 4. Specific heat measurements

Specific heat was measured using thermal relaxation technique performed with PPMS-9 (Quantum Design) in the temperature range 2–200 K. For the analysis of experimental data, the key although the most trouble issue is to separate the electron part containing information about superconducting properties, and lattice part of the specific heat. The reason is due the lattice part cannot be directly measured. To overcome the latter, it is possible to use so called corresponding states approximation [45]. In this approach, one turns to the lattice part of specific heat for a compound with a similar composition but undergoing neither superconducting nor magnetic transition. In case of Ba-122 family, we take the parent compound $BaFe_2As_2$ with a certain substitution and doping, while the parent compound undergoes a magnetic transition at 140 K. When varying the dopant or its concentration, the lattice parameters change by a few percent. To account this negligible variation, we use scaling coefficients close to unity. As an example, for $Ba_{0.67}K_{0.33}Fe_2As_2$ one may choose $Ba(Fe_{0.847}Co_{0.153})_2As_2$ [46], $Ba(Fe_{0.88}Mn_{0.12})_2As_2$ [47], $BaFe_{1.75}Ni_{0.25}As_2$ [48].

Mathematically, the corresponding states approximation could be expressed as follows:

$$C_{tot}^{SC}(T) = C_{exp}(T) = C_{e}^{SC}(T) + AC_{lat}^{nSC}(BT),$$

where $C_{tot}^{SC}(T)$ is whole specific heat corresponding to the experimental $C_{exp}(T)$, $C_{e}^{SC}(T)$ — electron part, $C_{lat}^{nSC}(T)$ is the lattice part for nonsuperconducting nonmagnetic compound, A and B — scaling coefficients. Above $T_C$, $C_{e}^{SC}(T)$ could be regarded as $\gamma_n T$. The A and B coefficients are determined using least square method considering the entropy conservation:

$$\int_0^{T_C} \frac{C_e}{T} dT = \int_0^{T_C} \gamma_n dT.$$

The specific heat of superconducting condensate could be calculated in framework of BCS model [49], however, for 122 compounds, it is preferable to describe the electron part using phenomenological two-band α-model [50]. This model elaborates specific heat of two-band superconductor as a sum of weighted partial contributions of the two condensates in each band. The fitting parameters are $\alpha_1 = 2\Delta_1/k_B T_c$, $\alpha_2 = 2\Delta_2/k_B T_c$ and $\varphi_1$ (where $\varphi_i = \gamma_i/\gamma_n$, $\gamma_i$ *is* the specific heat of i-th condensate in the normal state), which could be obtain using least square method.

The specific heat measurements were made with a 1.93 mg-piece of $Ba_{1-x}K_xFe_2As_2$ ($x = 0.33$) single crystal cleft from the same large crystal that was used for all other measurements; the sample possessed critical temperature $T_C = 36.5$ K. The raw experimental specific heat is shown in Fig. 3 at zero field. At temperatures tending to zero, the $C(T)/T$ dependence could be extrapolated to zero similarly to the Debye law $C(T)/T = \gamma(0) + \beta T^2$ [51], showing no features in the low temperature range (such as, e.g., growth towards the lowest T or Schottky anomaly), thus evidencing for high quality of the sample.

In the temperature interval 36–37 K the C(T) demonstrates a sharp peak related to the superconducting transition (see Fig. 3). The peak width is about 1 K, and the jump in the C/T data at the transition $\Delta C/T$ = 119 mJ/mol K$^2$. In order to separate the lattice and electron parts of, we used the specific heat of the reference compound Ba(Fe$_{0.88}$Mn$_{0.12}$)$_2$As$_2$ [31] since for the latter the specific heat data was measured within the widest temperature range. Figure 3a shows that the corresponding states approximation provides well agreement between the experimental data and those obtained with the modified lattice specific heat of Ba(Fe$_{0.88}$Mn$_{0.12}$)$_2$As$_2$. The resulting normalized electron part of the superconducting condensate $C_{es}/T\gamma_n$ for details, see [51]) fitted with various theoretical models is shown in the inset of Fig. 3. The single-band approach with an isotropic order parameter gives an optimum result for $2\Delta/k_BT_C$ = 3.7. Nonetheless, this approach is obviously insufficient to describe the superconducting properties of Ba-122: the corresponding theoretical C(T) (dashed line in the inset of Fig. 3) does not fit the remarkable hump in $C_{es}/T\gamma_n$ clearly seen at T/T$_C$ ~ 0.3–0.5. On the other hand, the phenomenological two-band approach (α-model, red line in the inset of Fig. 3) well reproduces the data. The difference between the model dependence and the experimental data does not exceed 5% of $C_{es}/T\gamma_n$, that corresponds to 4 mJ/molK$^2$. The deviation is within the measurements uncertainty and in relative units does not exceed 1% of the total measured $C_{exp}$. With the two band model, we find the following set of parameters: $\alpha_1 = 2\Delta_1/k_BT_C$ = 1.6 ± 0.1 ($\Delta_1$ = 2.5 ± 0.2 meV), $\alpha_2 = 2\Delta_2/k_BT_C$ = 7.2 ± 0.2 ($\Delta_2$ = 11.3 ± 0.3 meV), and $\varphi_1$ = 0.58 ± 0.02.

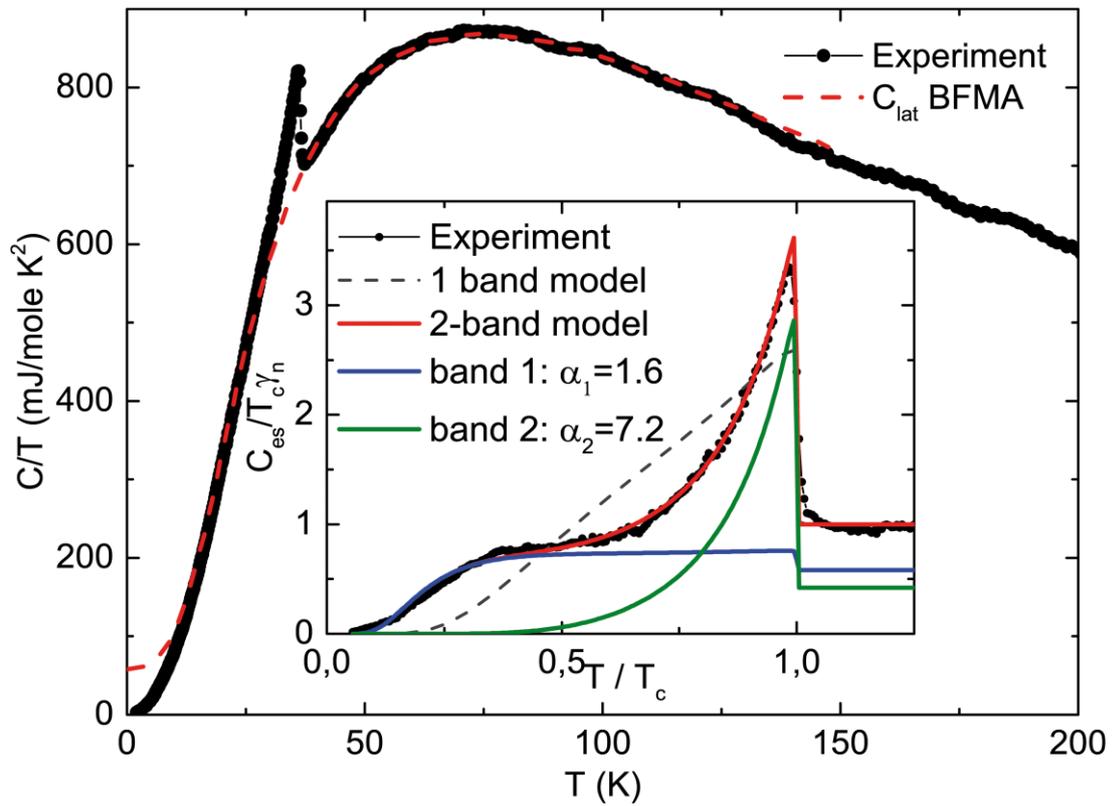

Fig. 3. a) Temperature dependence of the specific heat for $Ba_{0.67}K_{0.33}Fe_2As_2$ sample, normalized by temperature at zero field. Red dashed line the corresponding states approximation fit using the lattice specific heat of $Ba(Fe_{0.88}Mn_{0.12})_2As_2$ (BFMA). The inset shows the normalized electron part of the specific heat of the superconducting condensate $C_{es}/T\gamma_n$ fitted using single-band model (dashed line), and the two band BCS α-model (solid curves). The data are taken from [51].

## 5. Optical Spectroscopy

Optical spectroscopy is a basic technique to explore electrodynamic and superconducting properties [52,53]. The penetration depth is of the order of several hundreds of nanometers, thus facilitating the bulk probes, in particular, the measurements of superconducting gaps.

In case of single-band superconductor with an isotropic gap, electromagnetic radiation with the energy less than the superconducting gap value $2\Delta$, cannot be absorbed in the sample. This leads to the real part of the complex optical conductivity $\sigma$ tends to zero at $T \ll T_c$ and at frequency lower than that corresponding to the doubled superconducting gap $2\Delta$, herewith, the reflectivity coefficient tends to the unity. As a result, the optical response shows a feature at the frequency in the vicinity of $2\Delta$. In particular, for a bulk single crystal one should observe a peak in a spectrum of relative reflectivity $R(T)/R(T > T_c)$.

Here, due the $Ba_{1-x}K_xFe_2As_2$ ($x = 0.33$) sample was not large enough for accurate measurements of the absolute value of the reflection coefficient [51], we used the technique described in [54] to determine the superconducting gaps. This technique based on the measurement of relative reflection $R(T)/R(T > T_c)$ enables to minimize possible temperature-driven distortions of the optical setup, which may yield frequency dependent systematic errors in $R(\omega)$. It should be noted that for bulk s-wave superconductor the normalized reflectivity $R(T \ll T_C)/R_N$ (where $R_N$ is the reflectance in the normal state just above $T_C$) forms a maximum corresponding to $2\Delta$. For two-gap superconductor, the maximum is expected between the two SC gaps, closer to the one having a major contribution. The infrared-ranged measurements were done using Fourier IR reflection spectroscopy performed with Bruker Optics IFS-125HR spectrometer.

Figure 4 shows the normalized $R(T)/R(T = 40\ K)$ dependence measured at $T = 5–50\ K$. One can see that the normalized reflectivity $R(T)/R(40\ K)$ starts increasing as temperature $T$ decreases below $T_C$. This is because for s-wave superconductor at the temperatures below $T_C$ the reflectance approaches unity at energies $\hbar\omega < 2\Delta$. As a result, a peak at $\sim 160\ cm^{-1}$ (19.8 meV) correlates with the magnitude of the larger superconducting gap $\Delta_L \approx 10$ meV [25–27,47]. The smaller gap is beyond the frequency range of our IR measurements. The kink in the normalized reflectivity at $\sim 250\ cm^{-1}$ is probably caused by IR active phonon mode $E_u$ related to the Fe(ab)-As(−ab) vibrations [55]. This mode manifests itself in many $AFe_2As_2$ materials including A = Ca, Sr, Eu and Ba.

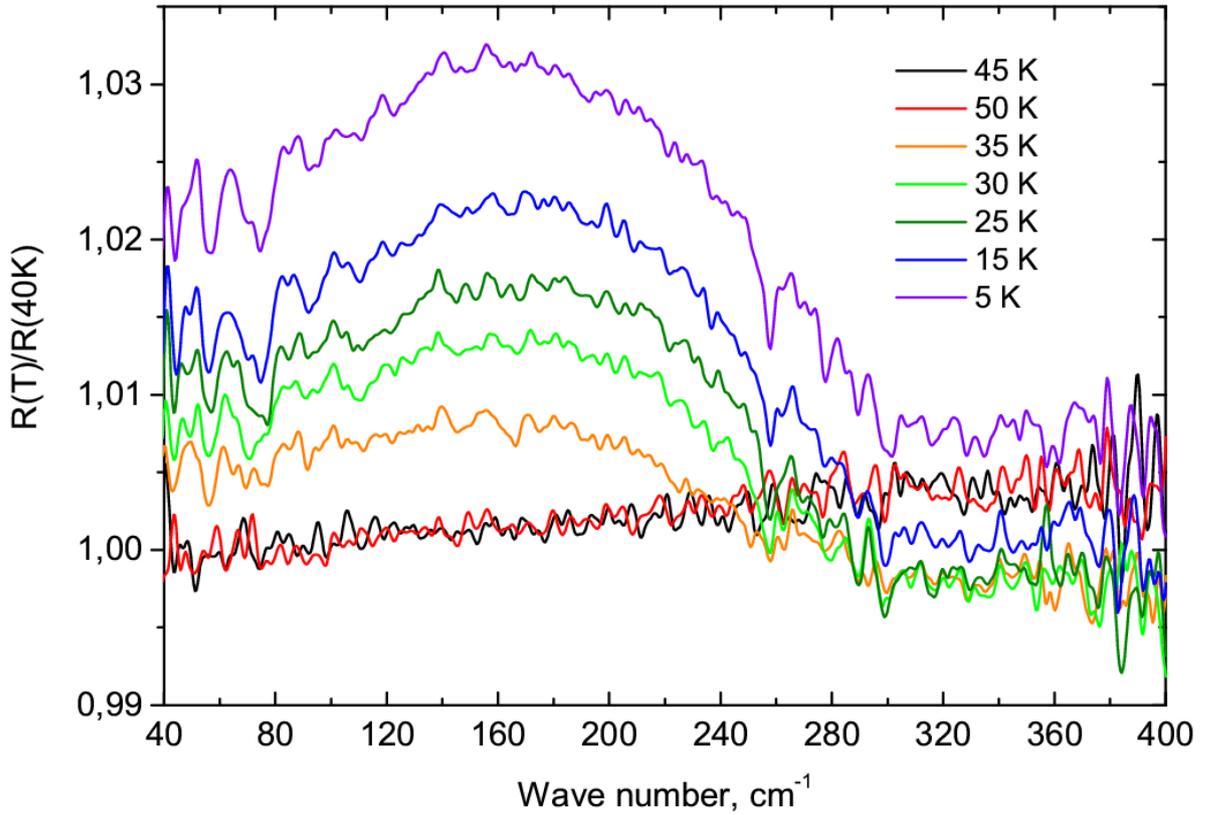

Fig 4. R(T)=R(40 K) dependences of $Ba_{0.67}K_{0.33}Fe_2As_2$ measured at T = 5–50 K. The data are taken from [51].

## 6. Intrinsic Multiple Andreev Reflection Effect (IMARE) Spectroscopy

Multiple Andreev reflections effect (MARE) spectroscopy is a unique direct probe of the bulk superconducting order parameter [56–58]. MARE occurs in ballistic [59] superconductor-normal metal-superconductor (SnS) junction which dimension $2a$ is less than the carrier mean free path $l$. MARE causes a pronounced excess conductance at low bias voltages (so called "foot"), and a subharmonic gap structure (SGS). In case of high-transparency (of the order of 95–98 %) SnS-junction, the SGS is a series of dynamic conductance dips at positions $V_n = 2\Delta/en$, where $\Delta$ is the superconducting gap, $e$ — elementary charge, and $n = 1, 2, \ldots$ — subharmonic order [60–63]. Probing SnS-junction gives a featured opportunity of direct determination of the gap value using the positions of Andreev subharmonics at $0 < T < T_C$ [60,63]. The latter enables one to directly measure temperature dependences of the gaps, and provides a local value of critical temperature (the $T_c$ corresponding to the contact area transition to a normal state). $T_c^{local}$ is essential for an accurate estimation of the most key superconducting parameter — a BCS-ratio $2\Delta/k_BT_c$. In two-gap superconductor, two distinct gaps would cause two SGS in the dI(V)/dV spectrum.

Angle-dependent in k-space superconducting gap strongly affects the shape of Andreev subharmonics [58,64,65]. Isotropic (*s*-wave) gap produces high-intensive and symmetrical dynamic conductance dips, while *d*-wave or fully anisotropic *s*-wave (having nodes at some angles in k-space) gap make the subharmonics poorly visible and strongly asymmetric. Anisotropic gap with $\cos(4\theta)$-type angle distribution in the $k_xk_y$-plane of the momentum space (which is very likely the case of Ba-122 [14,15]) causes doublet-like features in dI(V)/dV spectrum of tunneling contact for *c*-direction transport [66]. The doublet represents two minima connected by an arch, which positions correspond to the higher and lower extremes of the $\cos(4\theta)$-type gap angular distribution [58].

In Ba-122, high-quality SnS-contacts were formed by a "break-junction" technique [58,67]. The single crystal prepared as a thin rectangular plate was attached to a springy sample holder (oriented along the ab-plane) using four pads of In-Ga paste, and then cooled down to T = 4.2 K. In cryogenic environment, a gentle mechanical curving of the sample holder produced a cleavage of the sample, thus creating two superconducting banks separated with a weak link. An ScS-contact were formed, where *c* is a constriction. Judging by the resulting current-voltage characteristics (CVC) of the contacts made in Ba-122 [62], the constriction usually acts as a thin normal metal. The relation between the contact dimension and the carrier mean free path, the break junctions were in ballistic regime, thus making it possible to observe MARE [40,60–63,68]. In the set-up configuration, the superconducting banks slide apart each other along the ab-plane rather than separate at a valuable distance, therefore, our technique preserves the crack from impurity penetration and provides clean cryogenic surfaces to probe the gap(s) magnitude almost unaffected by surface defects. The easy mechanical readjustment also facilitates probing several tens of ScS contacts with various dimension, resistance, and transparency. This helps to collect a large amount of data with one and the same sample in order to check data reproducibility to be aware of dimensional effects.

Another unique feature typical for the "break-junction" barely, is the formation of natural ScSc-. . . -S arrays [40,51,56–58,68] in a layered sample. The layered single crystal usually exfoliates along the *ab*-planes with the formation of steps and terraces along the *c*-direction, where an intrinsic multiple Andreev reflection effect (IMARE) occurs. IMARE resembles an intrinsic Josephson effect [69] and was observed in cuprates, later in other layered superconductors (for a review, see [58]). When considering an Andreev array as a sequence of *m* identical SnS-junctions, the position of SGS scales with *m*: $V_n = 2\Delta \cdot m/en$, *n, m* =1*,* 2 *. . . .* In order to determine $m$ and the gap(s) value, one should find such natural numbers which scale the I(V) and dI(V)/dV curves for various arrays onto each other, or to achieve the same position of gap features with those for a single-junction spectrum. In array as in a natural structure of the crystal, a contribution of bulk effects well exceeds that of the surface influence [40,57,58]. Strictly speaking, the IMARE spectroscopy is currently the only technique probing the *bulk* values of superconducting gap(s) *locally* (within the contact area of 10–50 nm) [58]. The experimental set-up is detailed in [58,70].

Figure 5 shows CVC with excess current and a pronounced foot area at low bias voltages, which are footmarks of high-transparency SnS regime. The dynamic conductance spectra (data from [68]) correspond to the two SnS-contacts obtained with $Ba_{0.65}K_{0.35}Fe_2As_2$ single crystals from the same batch (the synthesis and characterization are detailed in [71,72]). The local critical temperature of these contacts is $T_C^{local} \approx 34$ K. The dynamic conductance spectra possess clear and well-reproduced doublet features at bias voltages corresponding to doubled large gap, $2\Delta_L \approx$ 12–16 meV. The lower spectrum also demonstrates the second subharmonics of the large gap at $V \approx \Delta_L/e \approx \pm(6-8)$ meV, where the range corresponds to the doublet edges. The width of the doublets is attributed to the in-plane anisotropy of the large gap in the momentum space. On the other hand, the external minima (which position determined the upper gap edge) are less pronounced as compared to those internal. This seems resulting from the more complex than $\Delta_L(\theta) \sim 0.5[1 + A \cos(4\theta)]$ angle distribution of the gap in k-space. The small gap SGS is resolved in the lower spectrum: the first at $V_1 \approx 3.4$ mV and the second at $V_2 \approx 1.7$ mV subharmonics are clear. A possible anisotropy of the small gap is an issue of further studies. A possible anisotropy of the small gap should be probed in further studies. Using the SGS expression, one easily get $\Delta_L \approx 6-8$ meV (~ 25% in-plane anisotropy), and $\Delta_S \approx 1.7$ meV. The inset shows temperature dependence of the upper extremum of the large gap.

In nearly optimal Ni-substituted crystals $BaFe_{2-x}Ni_xAs_2$ [40], we resolved the similar in-plane anisotropy of the large gap. In Figure 6, we compare the dI(V)/dV-spectra for two SnS arrays with various number of junctions formed by a sequent mechanical readjustment in one and the same sample [40]. The spectra look very similar, thus these contacts seem to be obtained in the same area of the cryogenic cleft. During the precise change in the holder curvature, the touching point for the two cryogenic clefts seemed to jump to a neighbour terrace, thus changing the number of acting layers from m = 10 (the upper spectrum) to m = 9 (the lower spectrum). In Fig. 6, the bias voltages of these curves were divided into these *integer m*, after the normalization, the position of the main dynamic conductance features coincide. The inset of Fig. 6 shows I(V) curves for these SnS arrays. A pronounced excess current near zero bias is typical for high-transparency Andreev mode. The contacts' resistance is large enough to provide a ballistic transport [59] and making it possible to observe IMARE. Black vertical bars point the n=1 and n=2 doublet SGS features for the large gap. Note that the n=2 doublet located at $V_2 \approx \pm(3.2–4.4)$ mV is twice narrower than the first one at $V_1 \approx \pm(6.5–8.8)$ mV corresponding to $2\Delta_L$, which is in a well agreement with the SGS formula. Therefore, the large gap is $\Delta_L \approx 3.2–4.4$ meV and has a ~ 30% in-plane anisotropy, similarly to that in the BKFA cited above [68]. Arrows in Fig. 6 point to the main (n=1) subharmonics for the small gap. These dips are much more intensive than those of the large gap, and do not match the expected position of the third subharmonics for $\Delta_L$ (expected at $V_3 \approx \pm(2.2–2.9)$ mV, according to the SGS formula). These doublets, although slightly overlap the $V_2$ position, determine the small gap $\Delta_S \approx 1.6$ meV.

To show unambiguously the features pointed by the arrows correspond to a distinct SGS, not related to the large gap, it is reasonable to probe the temperature influence to the spectrum. The dI(V)/dV at T = 4.2 K shown by the light green curve evolves at T = 8 K to that shown by the upper curve. Obviously, the arrow-pointed features significantly shift toward zero, accompanying with a dramatic reduction of their intensity with T increase by a factor of two. By contrast, it is not the case of the $\Delta_L$ subharmonics. Note that the dips at $V \approx \pm 1.2$ mV observed in both spectra do not correspond to the second subharmonic of the small gap and could be attributed to the beginning of the foot area at low biases.

The measure the dynamic conductance spectrum with temperature increase, in order to get direct temperature dependence of the gaps shown in Fig. 7. Noteworthy, the doublet structure of the $\Delta_L$ subharmonics is well resolved till the $T_c$, keeping the nearly constant 30 ± 3 % anisotropy (see Fig. 7b). The temperature dependences of outer and inner extremes of $\Delta_L$ are similar: the $\Delta_L(T)$ lies a bit lower than a single-gap BCS-like dependence (dash-dot line in Fig. 7). By contrast, the small gap behaviour differs: the $\Delta_S(T)$ is rather curved typically for an induced superconductivity within a wide range of temperatures caused by a proximity effect in k-space. Nonetheless, this feature of the $\Delta_S(T)$ is not a result of an induced $\Delta_L$ order parameter: obtained with the spectra of SnS-array, the temperature dependence delivers bulk properties of the material barely. The difference between the $\Delta_L(T)$ and $\Delta_S(T)$ temperature behaviour therefore points that the corresponding dI(V)/dV features relate to the two distinct superconducting condensates.

Remarkably, the experimental $\Delta_{L,S}(T)$ could be fit with a two-band model based on Moskalenko and Suhl system of equations with a renormalized BCS integral [73,74]. This system of equations determines a shape of gap temperature dependences using a set of electron-boson coupling constants $\lambda_{ij} = V_{ij}N_j$, where $i,j$ = L,S (hereafter the S index relates to the effective band with the small gap, L index — to that with the large gap), $V_{ij}$ are matrix interaction elements, $N^j$ is the normal density of states at the Fermi level. We take the average $\Delta_L(T)$ as a "driving" gap, the temperature dependence is shown in Fig. 7 by squares. In order to numerical fit of the experimental data using the BCS-like two-band model, we take the Debye energy $\hbar\omega_D = 20$ meV [75], and use the following fitting parameters: the ratio between the densities of states in the two bands $N_S/N_L$, and the intra- to the interband interaction ratio $\sqrt{V_L V_S}/V_{LS}$. The fitting with this model is detailed in [76,77]. Note only, the fitting cannot reveal the sign of interband constants $\lambda_{i\neq j}$, herewith, the obtained four constants are not full $\lambda_{ij}^{Full}$ (involving a Coulomb interaction $\mu^*$), but effective $\lambda_{ij} = \lambda_{ij}^{Full} - \mu^*_{ij}$.

The calculated theoretical curves $\Delta_{L,S}(T)$ (solid lines in Fig. 7) are typical for a case of a strong intraband and a moderate interband interaction. The deviation of the $\Delta_L(T)$ dependence from the single-band BCS-like curve seems resulting from an influence of the "weak" band having the larger density of states at the Fermi level. Due to nonzero interband coupling, both gaps turn to zero at one and the same critical temperature $T_c^{local}$.

Although the complex structure of the order parameter in Ba-122, the experimentally observed temperature dependence of the gaps well agrees with that predicted by the simple model. This enables one to use the estimated parameters in order to make some important conclusions concerning the superconducting state features of Ba-122. First of all, the "eigen" superconductivity (in a hypothetical case of zero interband interaction $V_{LS} = 0$) of the bands with the small gap tends to that described by the weak-coupling BCS-limit: in accordance with our estimations, the characteristic ratio $2\Delta_S/k_B T_c^S \approx 3.5$ (where $T_c^S$ is the "eigen" critical temperature of the $\Delta_S$-condensate at $V_{LS} = 0$). The set of estimated absolute values $\lambda_{LL} \approx 0.37$, $\lambda_{SS} \approx 0.23$, $|\lambda_{LS}| \approx 0.07$, $|\lambda_{SL}| \approx 0.02$, when supposing $\mu^* = 0$ typical for $s^\pm$-models [15,16] leads to extremely high density of states ratio $N_S/N_L \approx 3.5$, and a tiny interband coupling with $\sqrt{V_L V_S}/V_{LS} \approx 7.3$. On the other hand, when supposing a moderate Coulomb repulsion $\mu_{ij}^* = 0.13$, one gets the following set of the full constants: $\lambda_{LL} \approx 0.50$, $\lambda_{SS} \approx 0.36$, $|\lambda_{LS}| \approx 0.20$, $|\lambda_{SL}| \approx 0.15$. In the case, the density of states in the "weak" effective band is only 1.5 times larger than $N_L$ of the "driving" band, being in accordance with the band-structure calculations [78], and the intraband coupling is 2.7 times stronger than interband one. As follows, the moderate Coulomb repulsion is essential in order to describe the superconducting properties of iron-based pnictides.

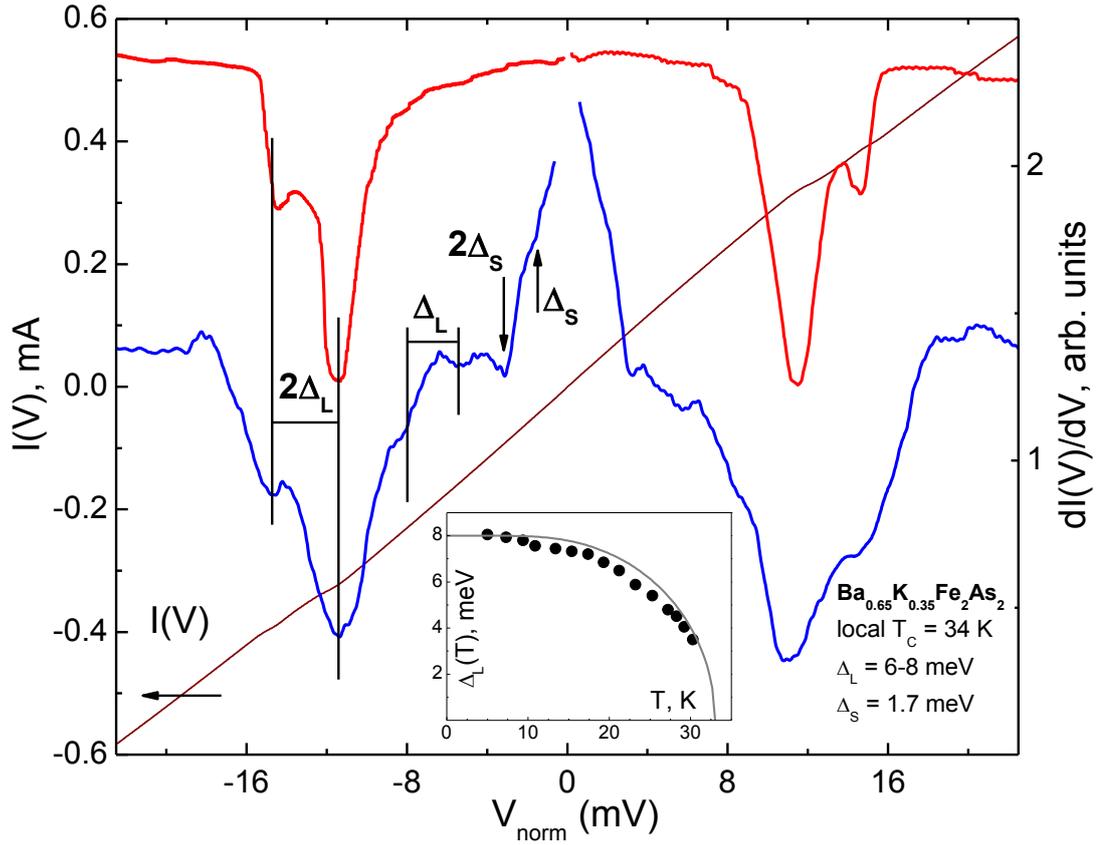

Fig. 5. Current-voltage characteristic I(V) (left vertical scale), and dynamic conductance spectra (right scale) for SnS-Andreev contacts with local critical temperature $T_C$ = 34 K measured in $Ba_{0.65}K_{0.35}Fe_2As_2$ single crystal at T = 4.2 K. The doublet features of the anisotropic large gap $\Delta_L$ = 6–8 meV (the range corresponds to the angle-dependent magnitude distribution in the k-space) and dips of the small gap $\Delta_S \approx 1.7$ meV are shown by vertical bars and arrows. The inset shows the large gap temperature dependence (circles), and single-band BCS-like behaviour for comparison. The data are taken from [51,68].

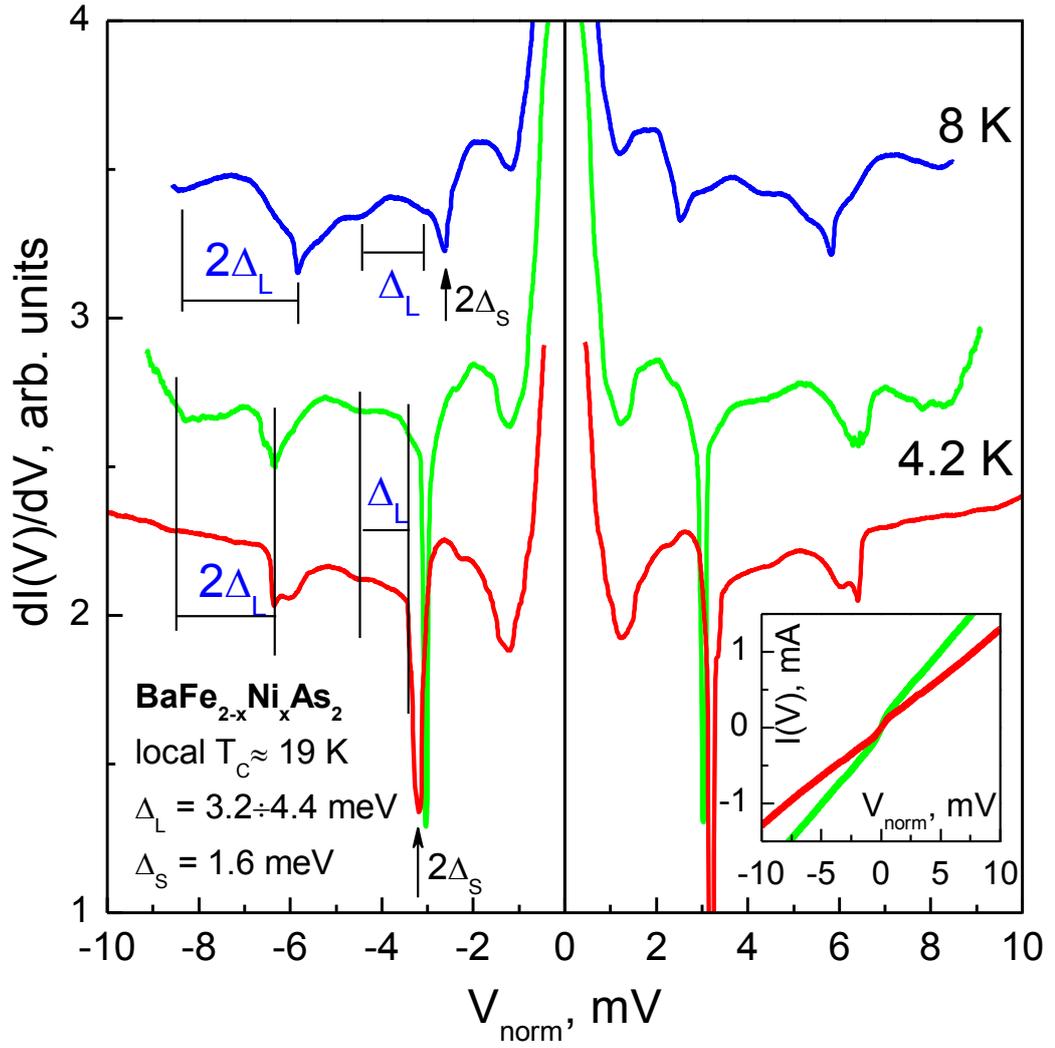

Fig. 6. Normalized dynamic conductance spectra for SnS-arrays with local critical temperature $T_C = 19$ K (the lower dI(V)/dV spectrum corresponds to m = 10 junctions in the array, the upper spectra — to m = 9 and T = 4.2 and 8 K) measured in BaFe$_{1.9}$Ni$_{0.1}$Fe$_2$As$_2$ single crystal. The doublet features of the anisotropic large gap $\Delta_L = 3.2–4.4$ meV and the small gap $\Delta_S = 1.6$ meV are shown by vertical bars and arrows. The inset shows the current-voltage characteristics of these contacts at T = 4.2 K. The data are taken from [40].

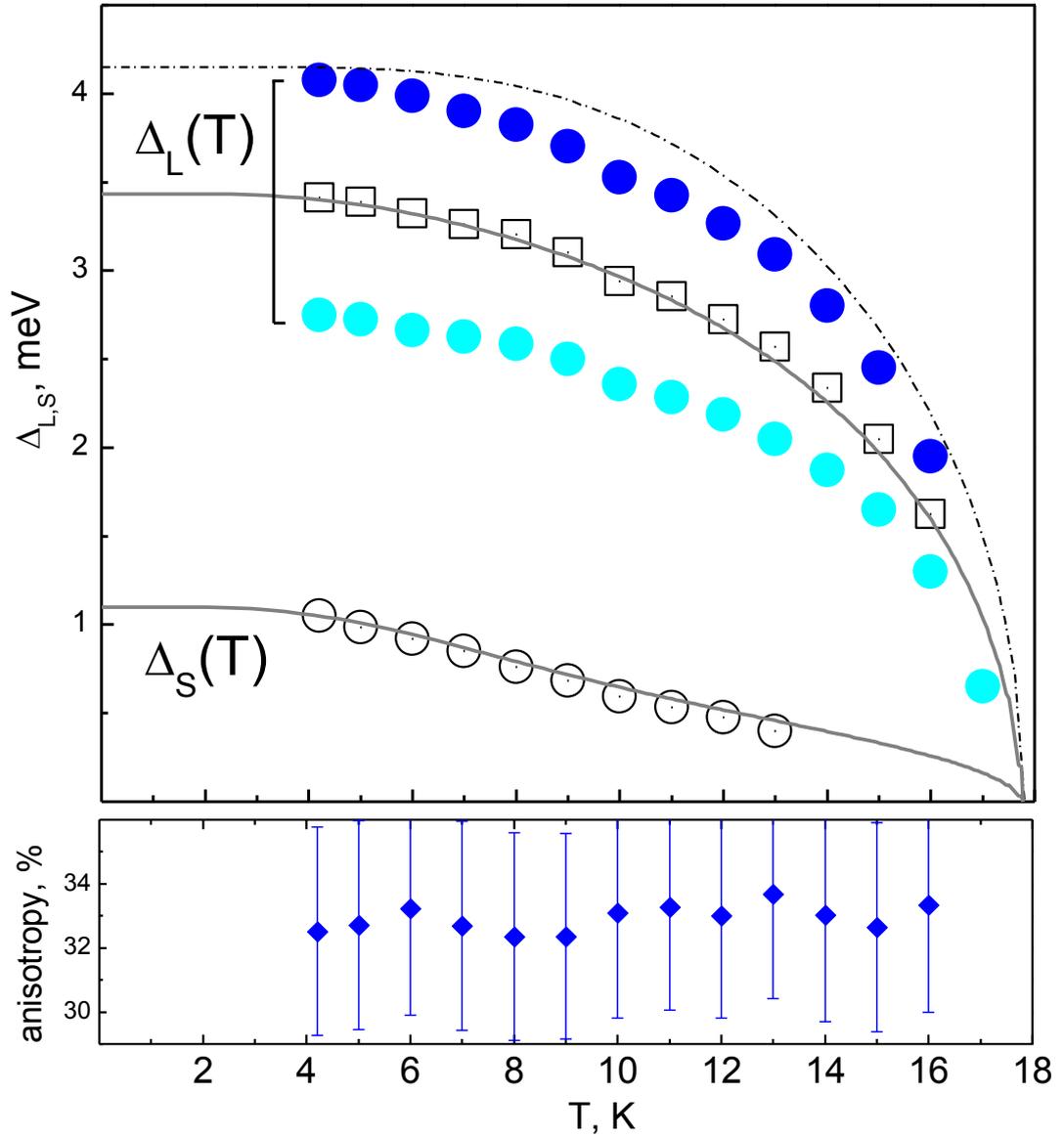

Fig. 7. a) Temperature dependences of the outer and inner extremes of the large gap (solid circles), and the small gap (open circles) in BaFe$_{1.9}$Ni$_{0.1}$Fe$_2$As$_2$. Squares depict the average $\Delta_L$ dependence. A BCS-like curve for $\Delta_L^{max}$ is shown by dash-dot line. Solid lines correspond to theoretical fits using two-band model based on Moskalenko and Suhl gap equations. b) The temperature variation of the large gap anisotropy taken as $1 - \Delta_L^{min}/\Delta_L^{max}$.

## 7. Measurements of the lower critical field

The measurement of the lower critical field technique using magnetization curves is based on the determination when the M(H) dependence become deviated from the linear M(H) ~ H behaviour, corresponding to vortex penetration into the bulk of the sample. The magnetization measurements were provided using squid-magnetometer MPMS-XL7 (Quantum Design).

In the vicinity of H$_{c1}$, the magnetization curve could be written as

$$M(H) = \begin{cases} aH + b, H < H^* \\ aH + b + c(H - H^*)^2, H > H^* \end{cases}$$

For all the determined $H^*$, at which magnetization was measured, we chose *a*, *b* and *c* parameters (*b* corresponds to a negligible deviation from zero of magnetization in zero field) to fit the experimental data. Then we calculate a correlation index, which demonstrates a clear maximum in the dependence on $H^*$; the maximum is located at $H_{c1}$.

Since in $Ba_{1-x}K_xFe_2As_2$ London penetration depth $\lambda \sim 100–200$ nm) is much larger than the coherence length $\xi \sim 2–2.5$ nm (for details, see [79]), a local London model is applicable to this compound. In the case, the normalized superconducting density is as follows:

$$\bar{\rho}_S^0(T) = \frac{\lambda_{ab}^2(0)}{\lambda_{ab}^2(T)} \approx \frac{H_{c1}(T)}{H_{c1}(0)}.$$

The resulting temperature dependence of the normalized superconducting density is shown in Fig. 8 [51] for $Ba_{0.67}K_{0.33}Fe_2As_2$ single crystal with $T_C \approx 37$ K.

Furthermore, in a case of single-band superconductor [80]

$$\bar{\rho}_S^0(T) = 1 + 2 \int_{\Delta(T)}^{\infty} \frac{\partial f}{\partial E} \frac{E dE}{\sqrt{E^2 - \Delta^2(T)}},$$

where $f = \exp[E/k_BT + 1]$ is the Fermi function, $\Delta(T)$ — a BCS-like gap temperature dependence, $E^2 = \varepsilon^2 + \Delta^2(T)$, E is an absolute energy, e is a single-particle energy offset the Fermi level. In this model, we used $H_{c1}(0)$ and $\alpha = 2\Delta(0)/k_BT_C$ as fitting parameters in order to reproduce the experimental data.

The lower critical field data in frames of single-band BCS-like model [80] is shown in Fig. 8. Obviously, the single-band model is insufficient to describe the data. Further, we used a phenomenological $\alpha$-model [80,81], where

$$\bar{\rho}_S^0(T) = \varphi \bar{\rho}_{S1}^0(T) + (1 - \varphi)\bar{\rho}_{S2}^0(T),$$

where $\bar{\rho}_{S1}^0(T)$ and $\bar{\rho}_{S2}^0(T)$ are the normalized superconducting densities for the two condensates taking with weight coefficients $\varphi$ and $(1 - \varphi)$. In this model, one should vary the four fitting parameters: $\alpha_1 = 2\Delta_1(0)/k_BT_C$, $\alpha_2 = 2\Delta_2(0)/k_BT_C$, the weight contribution of one band $\varphi$, and $H_{c1}(0)$. Obviously, the two-band model well reproduces the data points (see Fig. 8). We extract the following values of BCS ratios: $\Delta_L(0) = 11.5 \pm 0.5$ meV, $\Delta_S(0) = 2 \pm 0.35$ meV ($\varphi = 0.46 \pm 0.02$), and $2\Delta_L(0)/k_BT_C = 6.9 \pm 0.3$, $2\Delta_S(0)/k_BT_C = 1.2 \pm 0.2$. The $H_{c1}(0)$ is 25.5 Oe. Noteworthy, the $H_{c1}(0)$ is the lower critical field value determined with not accounted demagnetization factor for the certain sample. However, in the calculations normalized $H_{c1}(0)$ value is used, therefore, the shape of the temperature dependence is essential to obtain the superconducting parameters.

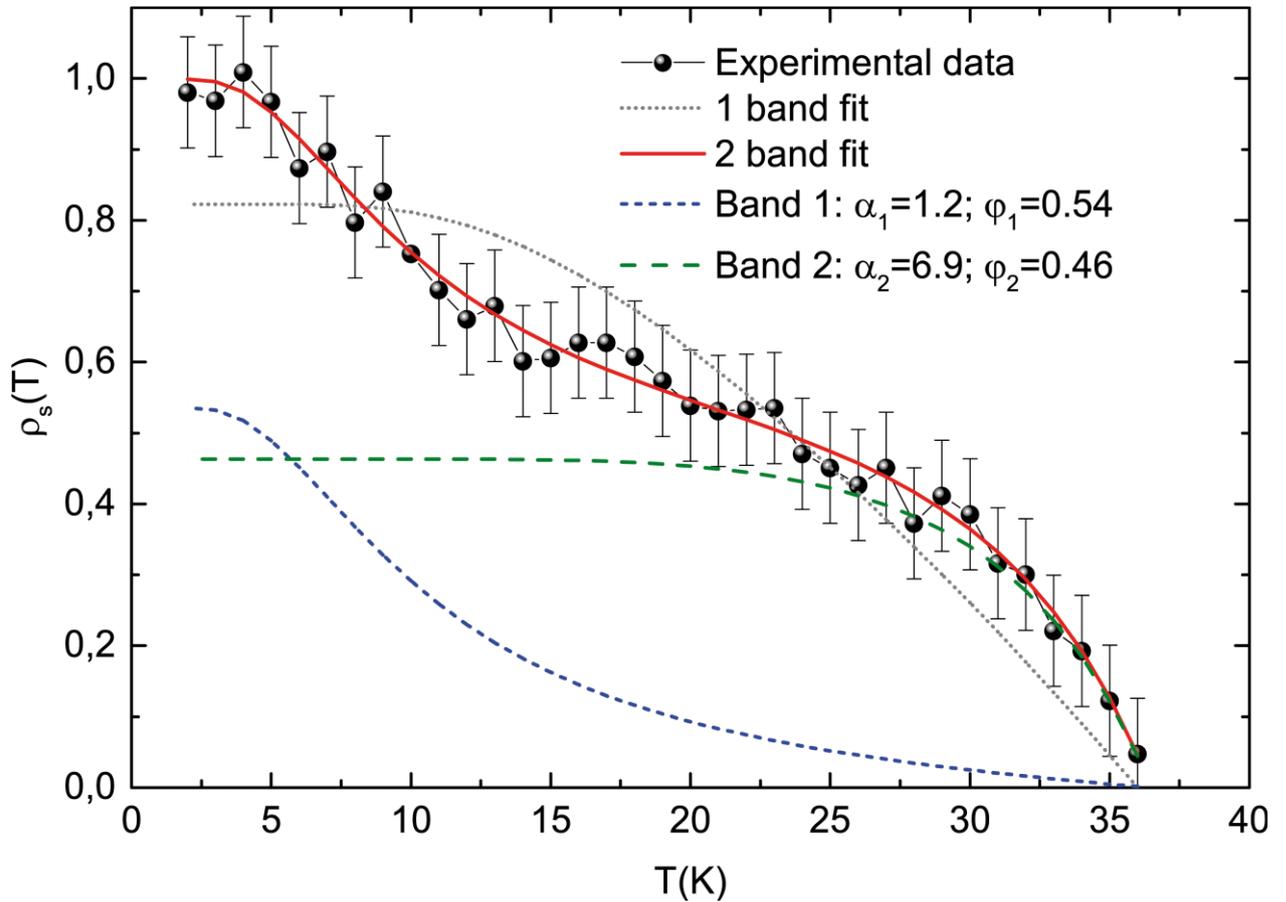

Fig. 8. Temperature dependence of the normalized superconducting density for $Ba_{0.67}K_{0.33}Fe_2As_2$ single crystal with $T_c \approx 37$ K fitted with a single-band (dashed line) and a two-band model (solid line). Partial contributions of the bands are shown by dotted lines. The data are taken from [51].

## 8. Discussion

Undoubtedly, the experimentally determined properties of strongly anisotropic compounds (in this case, having a layered crystal structure) a sensible to a lot of intrinsic and external influences, and to the experimental conditions. Contemporary stage of theoretical description of the multiple gap superconductors is not perfect yet. Obviously, comprehensive studies are necessary in order to explore the complex multigap structure of the order parameter in the Ba-122 family superconductors. Comparison between the results obtained by local and nonlocal, bulk and surface techniques is essential to get a reliable base to reveal the physics and features of the novel HTCS materials solely.

The data obtained in specific heat, lower critical field and SnS-Andreev spectroscopy studies revealed a presence of two distinct bulk superconducting order parameters. The large gap magnitude determined in optical probes is close to that integrated by the bulk as obtained in $H_{c1}(T)$, $C(T)$, and locally (MARE, IMARE). The bulk nature of the gaps is undoubted, since the $\Delta_L$ and $\Delta_S$ determined using single SnS-junctions (MARE) and natural Andreev arrays (IMARE) are reproducible and independent of the contact dimension or resistance. The temperature dependences $H_{c1}(T)$, $C(T)$, and $\Delta_{L,S}(T)$ well fitted with a two-band model also favor the latter. The majority of ARPES probes [19,20,22,23] confirm the two gaps with different values as well.

Fig. 9 shows the dependence of characteristic BCS-ratio versus Tc for the data obtained in our studies with $Ba_{1-x}K_xFe_2As_2$, $BaFe_{2-x}Ni_xAs_2$ compounds (red symbols), and the data from literature, including those for 122-arsenides with other composition. The in-plane gap anisotropy in k-space resolved in IMARE measurements is shown by red vertical bars. ARPES data (triangles) [19–24,27], lower critical field [32,68], specific heat [30,31], muon-spin-rotation (μSR) [23] and optical spectroscopy data [33] (squares) demonstrate a valuable diversity within the range $2\Delta_L/k_BT_C \approx 4.5- 7.5$. The following reasons could cause this contradiction:

a) local (tunneling techniques, Andreev spectroscopy, ARPES), and nonlocal ($H_{c1}(T)$, $C(T)$, IR-spectroscopy) probes of the order parameter, obviously, provide different $2\Delta_L/k_BT_C$ values in case of inhomogeneous sample;

b) the data by the bulk techniques are integrated throughout the entire sample and could be distorted due to a substantial energy dependence of the superconducting order parameter typical for strong-coupled superconductors. If it is the case, tunneling, Andreev and optical probes would reveal a so called gap edge $\Delta_{edge}$; on the other hand, bulk techniques give an averaged gap value, which could be "shifted" either lower or higher (depends on the $Re[\Delta(\omega)] > \Delta_{edge}$ to $Re[\Delta(\omega)] < \Delta_{edge}$ contribution rate). In our studies, the $2\Delta_L/k_BT_C$ determined in optical and Andreev studies are close, while those obtained using $H_{c1}(T)$ and $C(T)$ are a bit higher, thus favoring the latter;

c) the order parameter is possibly anisotropic in both $k_xk_y$-plane and $k_z$-direction, as discussed e.g. in [14]. In particular, a nonlinear $\Delta_L(k_z)$ dependence indirectly follows the periodic change of $\Delta_L$ value under the energy variation of emitted beam in ARPES studies [20]. In this case, the bulk techniques ($H_{c1}(T)$, $C(T)$) would give a gap value averaged over $k_z$-direction, thus different to that obtained by surface techniques (IR-spectroscopy, PCAR);

d) a possible disparity of the surface and bulk superconducting properties, which would distort the data by surface techniques;

e) a possible nontrivial in-plane angle distribution of the gap, another that $\Delta(\theta) \sim 0.5[1 + A \cos(4\theta)]$ (where $\theta$ is a $k_xk_y$-plane angle, $A < 1$) widely discussed in literature. Likewise, this could be a serious trouble when interpreting the gap features.

Nonetheless, our data obtained by five experimental techniques with Ba-122 compounds are in a good agreement:

1) Our data confirm the absence of nodes in the in-plane angle distribution of the large gap in nearly optimal (Ba,K)Fe$_2$As$_2$ with $T_c$ = 34–36.5 K, and in Ba(Fe,Ni)$_2$As$_2$ with $T_c \approx 18$ K;

2) The characteristic BCS-ratio $2\Delta_L/k_BT_C \approx 5.5$–7.2 well exceeding the BCS-limit 3.5 is a consequence of a strong coupling in the "driving" bands. For the small gap, the minor $2\Delta_S/k_BT_C \approx 1.2$–1.6 results from an induced superconductivity in these bands at $T > T_c^S$, where $T_c^S$ is the "eigen" critical temperature of the $\Delta_S$-bands and much less than the common $T_c$ of the compound. In the bands with the small gap, although their quasi-two-dimensionality, a weak superconductivity is developed, with an "eigen" BCS-ratio close to 3.5. Remarkably, in iron-based oxypnictides the intraband coupling within the small gap bands seems a bit stronger: according to our estimations, in average, $2\Delta_S/k_BT_c^S \approx 4$ [34,77,82]. Nonetheless, the weak superconductivity of $\Delta_S$-bands in Ba-122 is not exceptional: in magnesium diborides, "eigen" superconductivity in the 3D π-condensate similarly tends to the BCS-limit [76,77]. We believe a comparison of the properties of the "weak" bands in available two-gap superconductors, such as magnesium diborides and iron-based compounds, is a challenging issue, and, therefore, requires special theoretical studies.

3) The BCS-ratios determined in IMARE studies of (Ba,K) and (Fe,Ni) substituted single crystals within the wide range of critical temperatures, are in well agreement. The large gap order parameter scales within $T_c$ = 18–34 K: the $T_c$ change by a factor of 1.9 for (Ba,K)Fe$_2$As$_2$ causes nearly twice increase of the large gap, as compared with that in

Ba(Fe,Ni)$_2$As$_2$. Herewith, the anisotropy degree remains almost constant. Despite the electron doping by (Ba,K) substitution affects the structure of the spacer layers, while hole doping by (Fe,Ni) substitution directly distorts the superconducting blocks of the crystal lattice, one may conclude such the changes in composition do not seriously affect the underlying pairing mechanism in 122-arsenides. Similar scaling between $\Delta_{L,S}$ and $T_c$ was observed by us in iron-based oxypnictides of 1111 family, and 11-selenides [34]. An anisotropy of the small gap was not observed in our experiments, like in the majority of other studies. The only ARPES data are available [21] resolved the small gap anisotropy. Of course, this issue should be solved in further studies.

Despite the multi-orbital nature and a presence of at least three interacting bands at the Fermi level, the simple two-band model is enough to fit the temperature dependence of the most important parameters: the large and the small gap, electron specific heat, and the lower critical field. According to our estimations, two effective bands (where at $T < T_c$ the two condensates are developed with the gaps $\Delta_L$ and $\Delta_S$) interact rather weakly. The Tc value results mainly from a strong intraband coupling in the "driving" bands, with a nonzero Coulomb repulsion essential in order to describe the two-gap state of Ba-122 correctly. The latter points our experimental data seem to retreat the predictions of the initial $s^{\pm}$-model based on a strong interband pairing [15], thus favoring a realization of $s^{++}$-state [13,14].

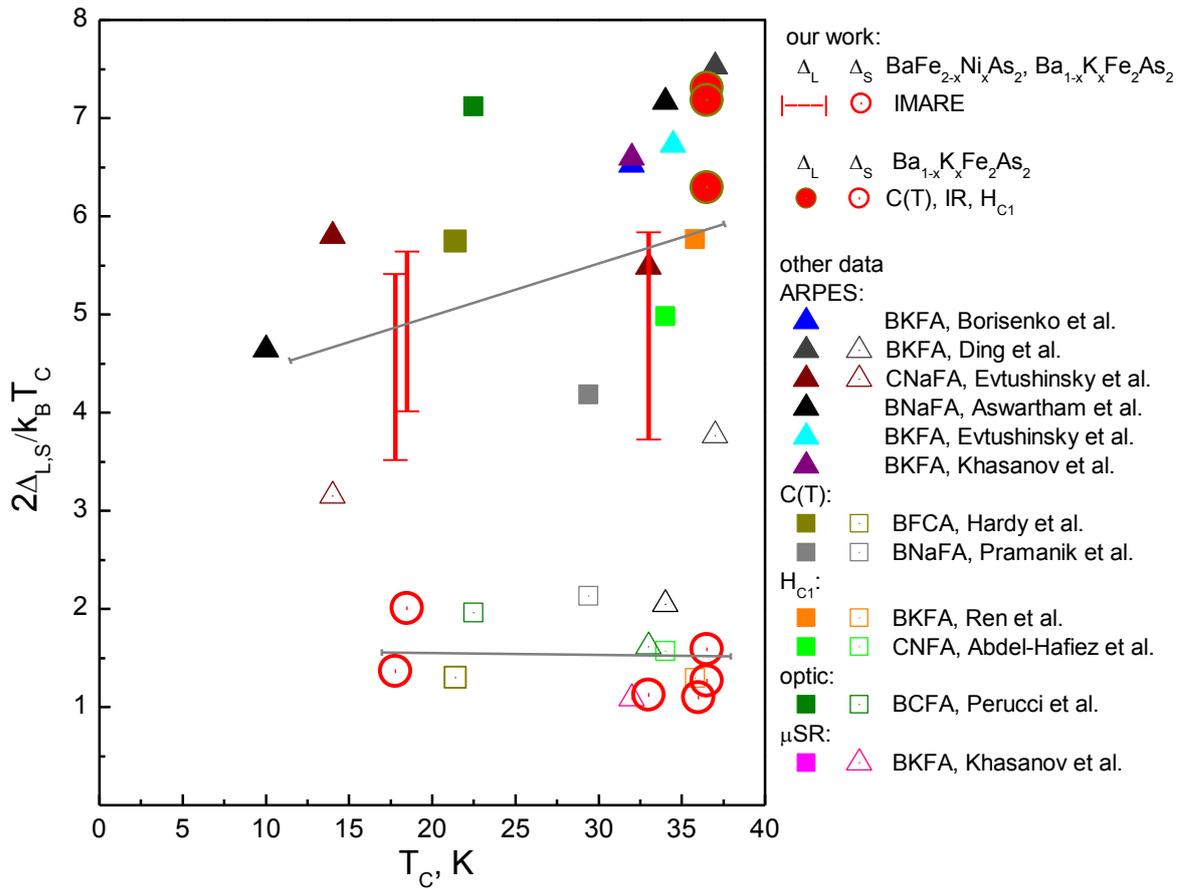

Fig. 9. The dependence of BCS ratio versus critical temperature for the large gap (solid symbols) and for the small gap (open symbols) for Ba-122 crystals with various compositions. Our data are shown by circles, the in-plane gap anisotropy resolved in IMARE probes is shown by red vertical bars. The ARPES data [7,19–24,27] (triangles), data using C(T), muon-spin-rotation, optical spectroscopy, and lower critical field measurements [30–33,68] (squares) are shown for comparison.

## 9. Conclusions

We present a comprehensive study of the $BaFe_{2-x}Ni_xAs_2$ и $Ba_{1-x}K_xFe_2As_2$ single crystals with hole and electron doping, correspondingly, belonging to the Ba-122 family of high-temperature superconductors. Despite the different type of dopant, both compounds demonstrate similar superconducting properties. The high-quality $Ba(Fe,Ni)_2As_2$ single crystals with various nickel concentration and critical temperatures up to $T_C \approx 21$ K were synthesized using self-flux method. The characterization revealed the presence of single superconducting phase and high homogeneity of the samples.

In order to study the structure of the superconducting order parameter, we used five complementary techniques.

The specific heat and lower critical field measurements gave information about the bulk properties, intrinsic multiple Andreev reflection effect (IMARE) spectroscopy was the direct local probe of the bulk superconducting parameters, whereas optical spectroscopy and ellipsometry probe crystal surface. Nonetheless, the results obtained using the various techniques are in good agreement. We mainly conclude the presence of two components of the superconducting condensate with unequal electron-boson interaction. The two gaps developed in separate Fermi surface sheets have no nodes in $k_xk_y$-plane and possess an extended s-wave symmetry, as agree with ARPES data.

The numerical data concerning the structure of the superconducting order parameter by the various techniques could be summarized as follows:

a) in optimal $Ba_{1-x}K_xFe_2As_2$ the large gap is $\Delta_L(0) = 8–11.3$ meV and has a ~ 30 % in-plane anisotropy; the small gap value is $\Delta_S(0) = 1.7–5=2.5$ meV;

b) for $Ba_{1-x}K_xFe_2As_2$ and $BaFe_{2-x}Ni_xAs_2$, the determined BCS-ratios $2\Delta_L/k_BT_c$ are close and well above the weak coupling limit due to a strong intraband electron-boson interaction in the bands with the large gap. The self-consistency of the order parameter structure seems a matter of the uniform pairing mechanism in these compounds, despite the different dopant type and $T_c$;

c) the large and the small gap, the electron specific heat, and the lower critical field decrease with temperature in the way another than that typical for single-band case. Remarkably, the two-band model seems sufficient to describe the most important superconducting parameters;

d) a moderate interband coupling, and the essentiality of a nonzero Coulomb repulsion when describing the two-gap superconducting state facilitate a possible realization of $s^{++}$-model.

**Acknowledgments**

We thank M.M. Korshunov, M. Abdel-Hafiez, Y. Chen, P.D. Grigoriev for the fruitful discussion and the provided materials. T.E.K., A.V.M., A.V.S., V.A.V., and Yu.F.E. acknowledge the Russian Science Foundation (project no. 16-12-10507) for the financial support. S.A.K., Yu.A.A., K.S.P., and V.M.P. acknowledge the Russian Science Foundation (project no. 16-42-01100) for the financial support. Yu.A.A. acknowledges the support of the Competitiveness Program of NRNU MEPhI.